\documentclass[%
aip,
adv,%
amsmath,amssymb,
preprint,%
]{revtex4-1}

\usepackage{sidecap}
\sidecaptionvpos{figure}{t}
\usepackage{graphicx}%
\usepackage{epstopdf}
\usepackage{color}
\usepackage{bm}%
\usepackage[colorlinks=true,linkcolor=blue,citecolor=blue,filecolor=blue,urlcolor=blue]{hyperref}
\usepackage{amsmath,amsfonts,amsgen,amsbsy,amsopn,amstext,amssymb}

\begin{document}

\title{Broadband quasi perfect absorption using chirped multi-layer porous materials}

\author{N. Jim\'enez}
\email{noe.jimenez@univ-lemans.fr}
\author{V. Romero-Garc\'ia}
\author{A. Cebrecos}
\affiliation{LUNAM Universit\'e, Universit\'e du Maine, CNRS, LAUM UMR 6613, Av. O. Messiaen, 72085 Le Mans, France}

\author{R. Pic\'o}
\author{V. J. S\'anchez-Morcillo}
 \affiliation{Instituto de Investigaci\'on para la Gesti\'on Integrada de zonas Costeras, Universitat Polit\`ecnica de Val\`encia, Paranimf 1, 46730, Grao de Gandia, Val\`encia, Spain}

\author{L. M. Garcia-Raffi}
 \affiliation{Instituto Universitario de Matem\'atica Pura y Aplicada, Universidad Polit\'ecnica de Valencia, Camino de Vera s/n, 46022, Valencia, Spain}

\begin{abstract}
This work theoretically analyzes the sound absorption properties of a chirped multi-layer porous material including transmission, in particular showing the broadband unidirectional absorption properties of the system. Using the combination of the impedance matching condition and the balance between the leakage and the intrinsic losses, the system is designed to have broadband unidirectional and quasi perfect absorption. The transfer and scattering matrix formalism, together with numerical simulations based on the finite element method are used to demonstrate the results showing excellent agreement between them. The proposed system allows to construct broadband sound absorbers with improved absorption in the low frequency regime using less amount of material than the complete bulk porous layer.
\end{abstract}

\maketitle


\section{Introduction}
\label{sec:intro}

Absorption is a major issue in acoustics and wave physics in general. The design of a perfect absorber is not a trivial matter since a twofold problem should be tackled. On one hand, the absorbent should present a perfect impedance matching in order to eliminate the sound reflection and, on the other hand, it should also have the amount of losses to absorb the energy of the wave without changing the impedance matching condition. More precisely, one can consider a perfect absorber as a lossy material that absorbs all the incoming waves (independent of its frequency and the impinging direction), whose impedance perfectly matches the surrounding media, avoiding the reflection. Perfect absorption (PA) is recently receiving an increasing interest \cite{Ma14,Ma16, Merkel15, Romero16a, Romero16b, Jimenez16}.

Chirped or graded materials are widely used in the wave physics community due to their possibilities to manipulate the wave propagation. These artificial materials are emerging as promising tools for potential applications in several branches of research and technology \cite{Chen14}. Several applications for focusing\cite{Martin10,Romero13a}, trapping \cite{Cassan11, Romero13, Zhu13}, bending waves \cite{Centeno06}, opening of wide full band gaps \cite{Kushwaha98} and controlling the spatial dispersion beams in reflection \cite{YuChieh14} have been developed. Recently, some of us presented a system with acoustic wave enhancement due to the progressive decrease of the group velocity along the propagation direction \cite{Romero13, Cebrecos14}. Chirped structures have also been used as efficiently absorbers. In the electromagnetic counterpart, an omnidirectional absorber has received a significant attention recently. It was shown that nearly total angle absorption of incident waves can be achieved using a cylindrical or a spherical device comprised of an absorbing core surrounded by a layer with dielectric constant varying with the distance $r$ to the device center \cite{Narimanov09}. The structure captures the incident radiation and guides it towards the absorbing core where the energy is dissipated. At the boundary between this layer and the absorbing core a nearly perfect impedance matching is achieved as well as at the air/layer interface, so the reflections are minimized. Acoustic analogues of this omnidirectional light absorber that works for airborne sound have been also proposed \cite{Climente12, Elliott14}.

In this work we focus on a 1D chirped multi-layer porous material. This system, contrary to the previous absorbing acoustic materials based on graded structures, does not has an absorbing core. By progressively varying the acoustical properties along the incident direction, the structure presents a quasi impedance matching condition that allows the penetration of the wave without reflection. As the waves propagate inside the material, the intrinsic viscothermal losses of the porous layers absorb the waves. The losses are tuned by changing the thickness of the porous layers, in such a way that the intrinsic losses compensate the leakage of the system, accordingly  to the critical coupling condition. This activates the perfect absorption.

We use the theoretical framework provided by the transfer and scattering matrix formulation in order to study the reflection and transmission properties of the system. By using the eigenvalues of the scattering matrix in the complex frequency plane \cite{Romero16a,Romero16b, Merkel15}, the intrinsic losses can be determined in order to compensate the leakage of the system to produce prefect absorption conditions. The chirped multi-layer porous material analyzed in this work behaves as a resonant Fabry-P\'erot cavity. Its resonances present a very low quality factor and they overlap to create a broadband unidirectional perfect absorber.

The work is organized as follows. Section \ref{sec:set-up} shows the setup and the main ingredients to create a chirp structure similar to the one we develop here. The theoretical framework based on the transfer and scattering matrices is shown in Section \ref{sec:theo}. Section \ref{sec:results} shows the main analytical results obtained in this work. The results presented in this work have been validated numerically using wave simulations based on a finite element approach of the 1D system, showing good agreement with the analytical results. The concluding remarks are shown in Section \ref{sec:conclusions}.

\section{Chirped multilayered porous structure} 
\label{sec:set-up}

\begin{figure}[t]
	\includegraphics[width=13.8cm]{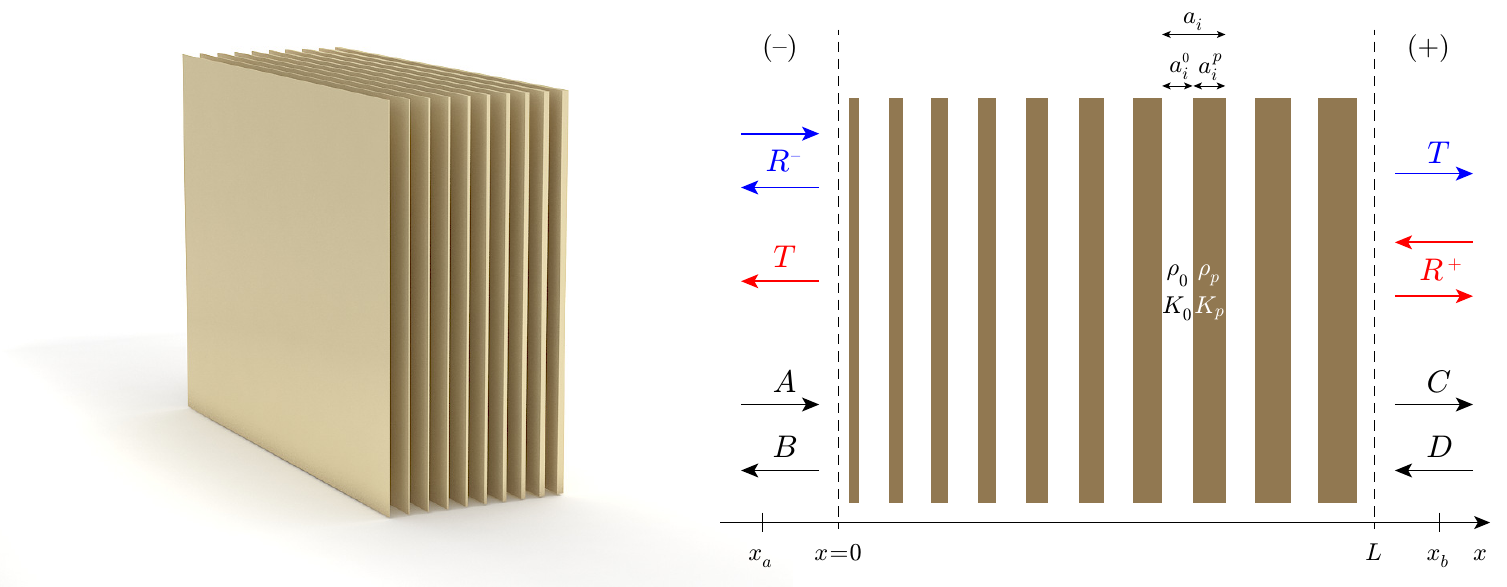}
	\caption{Scheme of the chirped multilayered porous structure. Shifted by $x^j=x^j-x_0^j$, to have the structure starting in $x^j_0=0$. Blue (Red) arrows represent the 1D reflection transmission problem produced by a plane wave radiating from the negative (positive) side of the structure. Black arrows represent the schematic representation for the amplitudes in the scattering matrix formulation.}
	\label{fig:fig1}
\end{figure}

In this work we deal with a discrete 1D structure made of $N$ layers (or unit cells). Each layer is formed by two sublayers of porous materials or air, see Fig.~\ref{fig:fig1}. The spatial distribution of porous and air sublayers is given by a linear chirped recurrence relation, 
$a_{i+1}^j = a_0^j - \mu^j\Delta x_i^j,$
with $i=1,...,N$ representing the layer number and $j=p,0$ the material being $p$ for the porous material and $0$ for air; $\mu^j$ is the adimensional chirp parameter, $a^j_i$ is the width of the $i$-th sublayer made of the material $j$, $a_0^j$ the width of the first sublayer of the chirp of material $J$, and $\Delta x_i^j$ is the total width of material $j$ used up to the $i$-th layer. Through this work, we consider the chirp direction in the positive $x$-axis, in such a way that if $\mu^j<0$ ($\mu^j>0$), the layer thickness is increased (decreased) in the chirp direction. The lattice constant of the $i$-th layer of the chirped structure is given by the addition of the width of the $i$-th sublayers of air and porous materials, $a_i=a_i^0+a_i^p$. The total length of the system is given by $L=\sum_{i=1}^Na_i$. This produces an asymmetric reciprocal structure. We consider a chirped structure produced by the recurrence relations with $a_0^p=\Delta x^p_1=1.9$ cm and $\mu^p=-0.2$ for the porous material and $a_0^0=\Delta x^0_1=3.0$ cm with $\mu^0=0$ for air. In this way, the sublayers of air have a constant width, $a_0^0$, and the porous sublayers increase their width all along the length of the structure, as shown in Fig.~\ref{fig:fig1}. 

These parameters have been obtained by nonlinear constrained optimization techniques \cite{powell1978} where the cost function to minimize was $e= \int_{f_1}^{f_2}{1-\alpha(f) df } $, with $f_1=20$ and $f_2=2000$ Hz, i.e. maximize the acoustic absorption of the system, $\alpha$, in the frequency range 20-2000 Hz. With these parameters, the structure made of $N = 10$ layers has a total length of $L = 34.9$ cm. It is worth noting here that the optimization of the geometrical parameters of the structure plays a fundamental role in the absorption of the system. On one hand, if the parameter $\mu^p$ excess from the optimal, the layer thickness grows abruptly and some reflection, $R$, is observed at the first layers due to impedance mismatch. On the other hand, if the parameter $\mu^p$ is lower than the optimal, the layer thickness grows slowly with distance. In this situation the structure can be phase matched at the first layers if the parameter $a_0^p$ is also low enough, i.e., in a structure composed by thin porous layers the reflection is absent. However, for a finite structure, transmission $T$, is observed due to the lack of losses at the end of the structure. To summarize, if the parameter $\mu^p$ is far from the optimal, some reflection or some transmission is produced and therefore, the absorption $\alpha= 1 - |R|^2 - |T|^2$ cannot be perfect.

\begin{table}[b]
\caption{\label{tab:tab1} Values of the physical parameters of the porous material.}
\begin{tabular}{ c c c c c }
\hline
$\phi_p$ & $\alpha_{\infty}$ & $\Lambda$ & $\Lambda'$ & $\sigma$ \\
 & & ($\mu$m) & ($\mu$m) & (Pa s)\\
\hline
\hline
 0.97 & 1.06 & 41 & 2$\Lambda$ & 51200\\
\hline
\end{tabular}
\end{table}

The porous material will be characterized by using the Johnson-Champoux-Allard (JCA) model \cite{johnson1987,champoux1991} that gives the expressions for the dynamic effective densities and bulk modulus of a porous material saturated by a fluid of density $\rho_0$ and bulk modulus $K_0$ considering a rigid frame. The porous material is characterized by its porosity, $\phi_p$, its tortuosity, $\alpha_{\infty}$, its flow resistivity, $\sigma$, and the thermal and viscous characteristic lengths, $\Lambda$ and $\Lambda'$ respectively. Table \ref{tab:tab1} gives the values of the static physical parameters of the porous material used in this work, which match those of a conventional rock-wool porous absorber\cite{olny2008}. The dynamic effective density and bulk modulus given by the JCA model are
\begin{eqnarray}
\rho_p &=& \alpha_{\infty}\rho_0\left[1 -\imath \frac{\sigma\phi_p}{\omega \rho_0 \alpha_{\infty}}\sqrt{1 + \imath \frac{4\alpha_{\infty}^2\eta\rho_0\omega}{\sigma^2\Lambda^2\phi_p^2}}~\right],\\
K_p  &=& \frac{\gamma P_0}{\gamma-(\gamma-1)\left[1 -\imath \frac{8\eta}{\Lambda'^2 \mathrm{Pr} \omega \rho_0 }\sqrt{1+ \imath\frac{\rho_0\omega \mathrm{Pr}\Lambda'^2}{16\eta}}~\right]^{-1}},
\end{eqnarray}
where $\imath=\sqrt{-1}$ and $\omega$ the angular frequency. Considering that the saturating fluid is air, $\rho_0=1.213$ kg m$^{-3}$, $\mathrm{Pr}=0.71$ is the Prandtl number, $\gamma=1.4$ is the ratio of the specific heats, $P_0=101325$ Pa is the atmospheric pressure and $\eta=1.839\;10^{-5}$ kg m$^{-1}$s$^{-1}$ is the dynamic viscosity. It is worth noting here that the sound velocity in air is given by $c_0=\sqrt{{\gamma P_0}/{\rho_0}}$. With these expressions one can obtain both the effective wavenumber in the porous material, by using $k_p = {\omega}/{c_p}=\omega\sqrt{\rho_p/K_p}$, where $c_p$ is the effective sound speed in the porous material, and the acoustic impedance, $Z_p = \sqrt{\rho_p K_p}$.

\section{Transfer and scattering matrices} 
\label{sec:theo}

The system described before can be analyzed either by using the transfer matrix or by using the scattering matrix. The two matrices are related and can give directly the reflection and transmission coefficients of the analyzed materials as well as its effective parameters among other properties. In this Section we briefly present the two methods and the relations between them.

The transfer matrix between the two faces of the homogeneous and isotropic material $j$, $\mathbf{t}^j$, extending from $x=0$ to certain thickness $x=l$, is used to relate the sound pressure, $P$, and normal acoustic particle velocity, $V$, between its two faces, i.e.,
\begin{eqnarray}
\left[\begin{tabular}{c}
$P$\\
$V$
\end{tabular}\right]_{x=0}=
\mathbf{t}^j
\left[\begin{tabular}{c}
$P$\\
$V$
\end{tabular}\right]_{x=l}=
\left[\begin{tabular}{cc}
$t^{j}_{11}$ & $t^{j}_{12}$\\
$t^{j}_{21}$ & $t^{j}_{22}$
\end{tabular}\right]
\left[\begin{tabular}{c}
$P$\\
$V$
\end{tabular}\right]_{x=l}
=
\left[\begin{tabular}{cc}
$\cos(k_jl)$ & $\imath Z_j\sin(k_jl)$\\
$\imath \sin(k_jl)/Z_j$ & $\cos(k_jl)$
\end{tabular}\right]
\left[\begin{tabular}{c}
$P$\\
$V$
\end{tabular}\right]_{x=l},\nonumber\\
\end{eqnarray}
where $j=p,0$ in our system (see Fig.~\ref{fig:fig1}). Then the transfer matrix of the $i$-th layer, $\mathbf{T}^i$, that is formed by two sublayers of air and porous material can be defined as follows,
\begin{eqnarray}
\mathbf{T}^{i}=\left[\begin{tabular}{cc}
$\cos(\frac{k_0a^0_i}{2})$ & $\imath Z_0\sin(\frac{k_0a^0_i}{2})$\\
$\imath \frac{\sin(\frac{k_0a^0_i}{2})}{Z_0}$ & $\cos(\frac{k_0a^0_i}{2})$
\end{tabular}\right]
\left[\begin{tabular}{cc}
$\cos(k_pa^p_i)$ & $\imath Z_p\sin(k_pa^p_i)$\\ 
$\imath \frac{\sin(k_pa^p_i)}{Z_p}$ & $\cos(k_pa^p_i)$
\end{tabular}\right]
\left[\begin{tabular}{cc}
$\cos(\frac{k_0a^0_i}{2})$ & $\imath Z_0\sin(\frac{k_0a^0_i}{2})$\\
$\imath \frac{\sin(\frac{k_0a^0_i}{2})}{Z_0}$ & $\cos(\frac{k_0a^0_i}{2})$
\end{tabular}\right].\nonumber\\ 
\end{eqnarray} 
Therefore the total transfer matrix of the whole system, $\bf{T}$, can be obtained by the product of the transfer matrices of each layer of the chirped material. Thus, the total transfer matrix is given by
\begin{eqnarray}
\mathbf{T}&=&
\left[\begin{tabular}{cc}
$T_{11}$ & $T_{12}$\\
$T_{21}$ & $T_{22}$
\end{tabular}\right]
=
\prod_{i=1}^N \mathbf{T}^i .
\label{eq:totalmatrix}
\end{eqnarray}

The reflection and transmission coefficients can be directly calculated from the elements of the matrix given in Eq. (\ref{eq:totalmatrix}) as
\begin{eqnarray}
\label{eq:T}
T&=&\frac{2e^{\imath k L}}{T_{11}+T_{12}/Z_0+Z_0T_{21}+T_{22}},\\
\label{eq:Rm}
R^-&=&\frac{T_{11}+T_{12}/Z_0-Z_0T_{21}-T_{22}}{T_{11}+T_{12}/Z_0+Z_0T_{21}+T_{22}},\\
\label{eq:Rp}
R^+&=&\frac{-T_{11}+T_{12}/Z_0-Z_0T_{21}+T_{22}}{T_{11}+T_{12}/Z_0+Z_0T_{21}+T_{22}},
\end{eqnarray}
where the superscripts $(+,-)$ denote the incidence direction, i.e., the positive and negative $x$-axis incidence respectively (see Fig.~\ref{fig:fig1}). We notice that, if the system was symmetric, then, $T_{11}=T_{22}$, and as a consequence, $R^+=R^-$. However, this property is not retained by the asymmetric chirped structure. The reciprocal behaviour of the system can be seen from the fact that the transfer matrix is unitary, i.e., its determinant is one ($T_{11}T_{22}-T_{12}T_{21}=1$).

On the other hand, the scattering matrix, $\bf{S}$, relates the amplitudes of the incoming waves to the system with those of the out-coming waves. The $\bf{S}$-matrix is widely used in wave physics to characterize the wave scattering. The poles and zeros of the $\bf{S}$-matrix in the complex-frequency plane provide very rich information as they are identified with bound states, virtual states or resonances.

We consider that the total pressure in a point $x_a<0$ ($x_b>L$) is given by $p(x_a)=Ae^{-\imath k x_a}+Be^{\imath k x_a}$ [$p(x_b)=Ce^{-\imath k x_b}+De^{\imath k x_b}$]. Then the relation between the amplitudes is given by the scattering matrix as
\begin{eqnarray}
\label{eq:S}
\left[\begin{tabular}{c}
$A$\\
$D$
\end{tabular}\right]=
\left[\begin{tabular}{cc}
$S_{11}$ & $S_{12}$\\
$S_{21}$ & $S_{22}$
\end{tabular}\right]
\left[\begin{tabular}{c}
$C$\\
$B$
\end{tabular}\right]
=
\left[\begin{tabular}{cc}
$T$ & $R^-$\\
$R^+$ & $T$
\end{tabular}\right]
\left[\begin{tabular}{c}
$C$\\
$B$
\end{tabular}\right].
\end{eqnarray}
The relation between \textbf{T} and \textbf{S} is then given by the Eqs.(\ref{eq:T}-\ref{eq:S}).

In order to calculate the effective parameters of the $i$-th layer, we consider that each layer can be characterized as a slab of effective material with effective wavenumber, $k^i_{\mathrm{eff}}$, and effective impedance, $Z^i_{\mathrm{eff}}$. Then,
\begin{eqnarray}
\left[\begin{tabular}{cc}
$T^i_{11}$ & $T^i_{12}$\\
$T^i_{21}$ & $T^i_{22}$
\end{tabular}\right]
=\left[\begin{tabular}{cc}
$\cos(k^i_{\mathrm{eff}}L)$ & $\imath Z^i_{\mathrm{eff}}\sin(k_{\mathrm{eff}}L)$\\
${\imath \sin(k^i_{\mathrm{eff}}L)}/{Z^i_{\mathrm{eff}}}$ & $\cos(k^i_{\mathrm{eff}}L)$
\end{tabular}\right].
\end{eqnarray}
The effective parameters of the $i$-th layer can be obtained from its transfer matrix as follows
\begin{eqnarray}
	k^i_{\mathrm{eff}}=\frac{1}{L}\cos^{-1}\left(\frac{T^i_{11}+T^i_{22}}{2}\right) \quad  \mathrm{,}\quad
Z^i_{\mathrm{eff}}=\sqrt{\frac{T^i_{12}}{T^i_{21}}}.
\end{eqnarray}


\section{Results}
\label{sec:results}
\begin{figure}[tbp]
	\includegraphics[width=13.9cm]{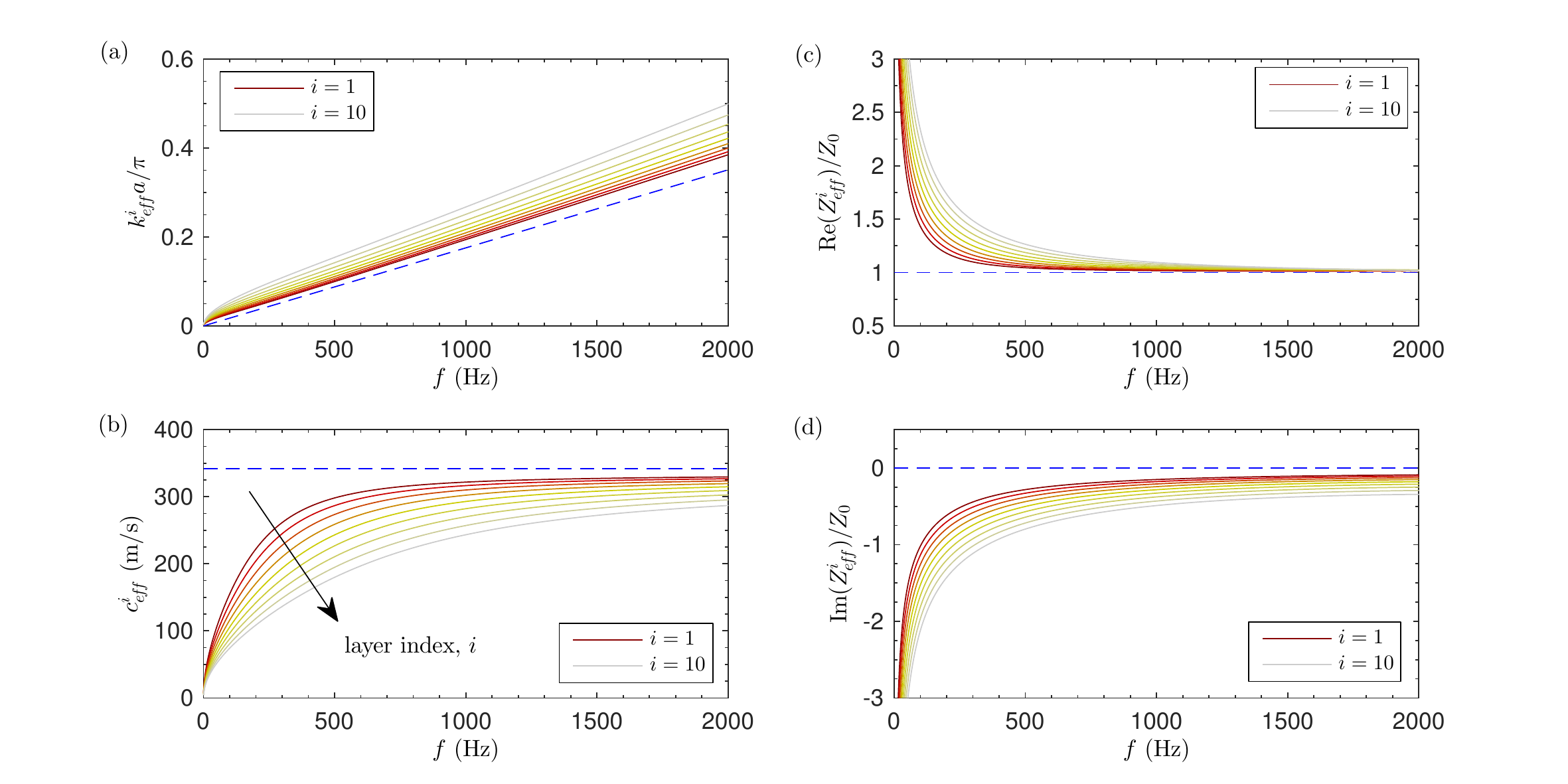}
	\caption{Analysis of the effective physical parameters. (a) Effective normalized wave number in terms of frequency for each layer of the system. (b) Effective sound speed in terms of frequency for each layer. (c) and (d) the real and imaginary parts of the normalized acoustic impedance for each layer, respectively. In all the cases the reference values of air are represented by the dashed blue lines.}
	\label{fig:fig2}
\end{figure}

Using the transfer and matrix formalisms presented above, we analyze the wave propagation trough the chirped multi-layer porous material previously introduced. In order to validate the results we use a numerical approach based on the Finite Element Method (FEM). For solving the problem using FEM, it is necessary to define the symmetry, discretize the domain and consider the boundary conditions. In the boundary of each layer both the continuity of the pressure and the velocity are imposed, the properties of the porous media come from the JCA model as previously mentioned. Plane waves radiate the structure and perfectly matched layers (PML) are used to simulate the Sommerfeld conditions in the bounds of the numerical domain. In our simulations the number of degrees of freedom was 2069 and the computational time for the whole range of frequencies was 20 s.

\subsection{Local parameters}
\label{sec:matching}
We start by analyzing the wave propagation inside the chirped multi-layer material. The chirped distribution allows the system to have a soft variation of the physical properties all along the chirped direction, in such a way that the local physical properties at the $i$-th layer, can be considered as those of an infinite system made of the periodic distribution of the $i$-th layer.

Figure~\ref{fig:fig2} shows the frequency dependent effective parameters for each layer in the chirped multi-layer system. The properties of the first layer, i.e., $k_{\mathrm{eff}}^1$, $c_{\mathrm{eff}}^1$ and $Z_{\mathrm{eff}}^1/Z_0$, are close to ones of the surrounding medium (air, dashed lines) for a broadband range of frequencies starting from 500 Hz. This results in a quasi-matching impedance condition at the entrance of the system for a whole range of frequencies. However as the wave propagates inside the multi-layer system, the local properties adiabatically change producing slow sound conditions (decreasing value of the effective velocity all along the structure) and introducing losses (increasing of the real part of the impedance all along the structure) due to the increasing amount of the porous material in each layer. At the end of the system, there is a mismatch condition but, as we will see later, the losses in the structure are high enough to efficiently absorb the acoustic waves in the direction of the chirp for a broadband range of frequencies avoiding reflection of waves. 


\subsection{Impedance matching condition}
\label{sec:reflection}

\begin{figure}[htbp]
	\centering
	\includegraphics[width=13.9cm]{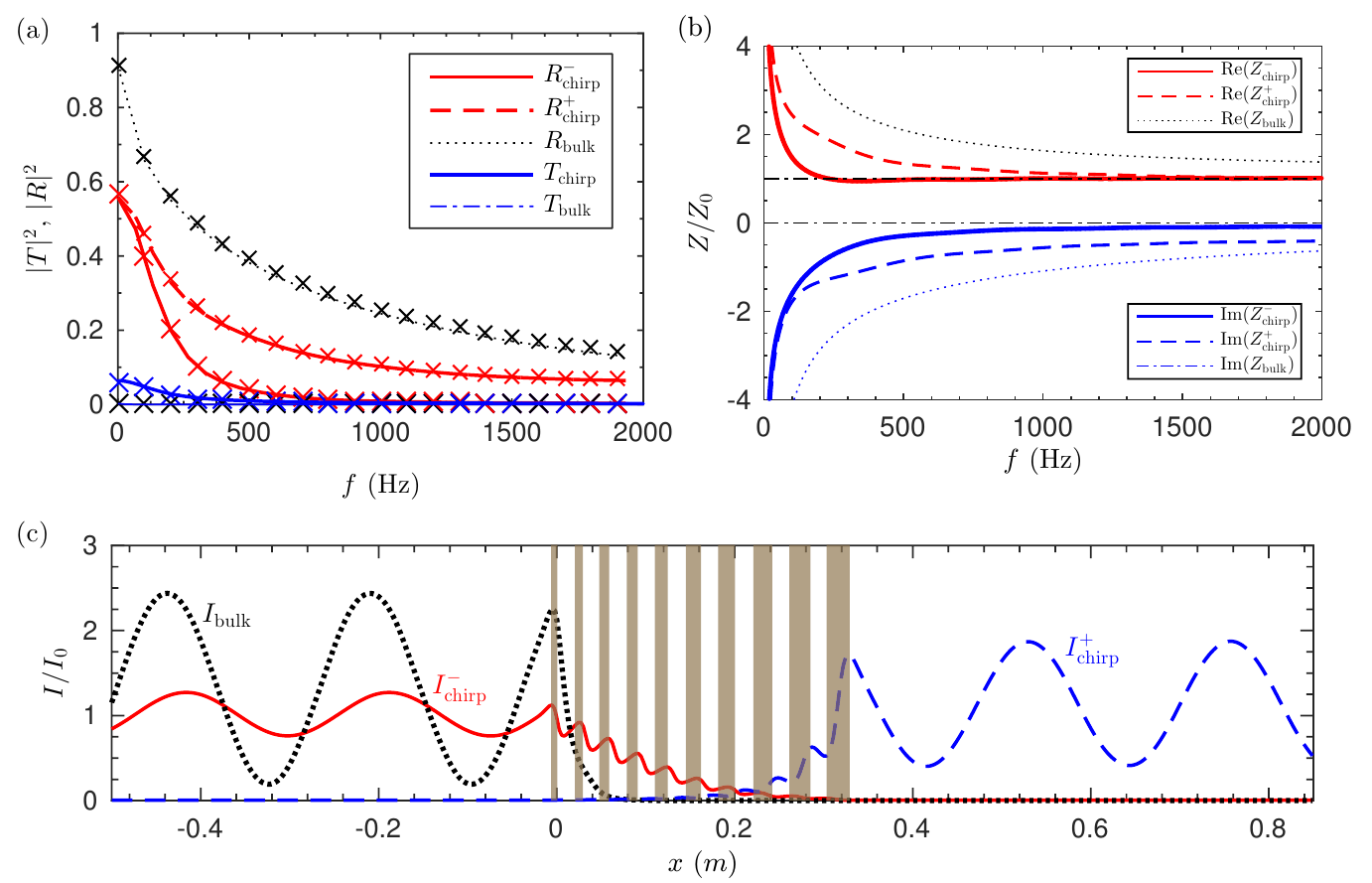}
\caption{(a) Reflection and Transmission coefficients. Red thick continuous (dashed) line represents the reflection coefficient, $|R_\mathrm{chirp}^-|^2$ ($|R_\mathrm{chirp}^+|^2$), of the chirped structure along the (opposite) direction of the chirp. Black thin dotted line represents the reflection coefficient of a bulk porous material with the same thickness and material properties than the chirped multi-layer structure. Blue thick line represent the transmission, $|T_\mathrm{chirp}|^2$, through the chirped and blue dotted-dashed the bulky porous material. In all the cases the markers ($\times$) represent the numerical results from the FEM model. (b) Acoustic impedances of the whole system. Red (blue) lines represent the real (imaginary) part of the total impedance of the system. Continuous lines show impedances for the chirped structure for the ($-$) incident direction, dashed lines the positive (+) direction and dotted lines the impedance of the bulk porous material. (c) Acoustic intensity along the system from both sides at 750 Hz. Red continuous (blue) line represent the intensity profile due to an plane wave incidence along (opposite) the chirp direction. Black dotted line shows the acoustic intensity of the bulk porous material with the same dimensions as the chirp. Vertical shaded areas show the position of the porous sublayers of the chirp.}
\label{fig:fig3}
\end{figure}

Figure~\ref{fig:fig3} shows the analysis of the reflection and transmission problems.The results obtained using the transmission matrix model are in perfect agreement with the results obtained using the FEM simulations. Figure~\ref{fig:fig3}~(a) shows the reflection and transmission coefficients of the chirped structure along the two incidence directions and those of the bulky porous material with the same thickness, $L$, and material properties than the chirped multi-layer structure. First of all we notice that, the transmission coefficients from both sides of the chirped structure are identical. However, and as a consequence of the asymmetry of the problem, we can see differences between the values for the reflection coefficient from each side. On the other hand, the reflection coefficient from the negative side of the structure presents very low values. This effect can be explained by analyzing the impedance of the whole system.

The acoustic impedance from each side of the system can be reconstructed from the reflection and transmission coefficients as follows\cite{Markos03,Chen04}
\begin{eqnarray}
Z^{[+,-]}_\mathrm{chirp}=Z_0\sqrt{\frac{\left(1+R^{[+,-]}_\mathrm{chirp}\right)^2-T_\mathrm{chirp}^2}{\left(R^{[+,-]}_\mathrm{chirp}-1\right)^2-T_\mathrm{chirp}^2}}.
\end{eqnarray}
Figure~\ref{fig:fig3}~(b) shows the impedances of the chirped and the bulk material. Results clearly show that the chirped structure is closer to the impedance matching condition ($Z/Z_0=1+0\imath$) than the bulk material above 500 Hz. This will explain, first of all, that the reflection coefficients from both sides of the chirped structure are smaller than that of the bulk material.Moreover, if one pay attention to the impedances from each side of the structure, the impedance from the chirped direction has an imaginary part, $\mathrm{Im}(Z^{-}_\mathrm{chirp}$), closer to zero, i.e., a smaller reactance than along the opposite direction. Also, the real part of the impedance from the chirped direction, $\mathrm{Re}(Z^{-}_\mathrm{chirp}$), has a value closer to one than the corresponding along the opposite direction. This combined behavior of the real and imaginary parts of $Z^{-}_\mathrm{chirp}$ for a broadband range of frequencies (above 500 Hz), produces a quasi impedance broadband matching condition along the chirped direction, so soft reflections from the negative part, i.e., near zero values of $R_\mathrm{chirp}^-$.

In order to observe the effect of the acoustic field of these soft reflections, Fig.~\ref{fig:fig3}~(c) shows the acoustic intensity profile for the different sides at 750 Hz. At this frequency, $|R^-_\mathrm{chirp}|^2=0.015$, $|R^+_\mathrm{chirp}|^2=0.13$, $|R_\mathrm{bulk}|^2=0.3$. The impedance condition in the negative part of the structure produces a soft oscillation around $I/I_0=1$ for the case of the chirped structure excited with a plane wave coming from the negative part. Contrary, the oscillations of the intensity for the other two cases are more pronounced due the impedance mismatching.

\subsection{Unidirectional quasi perfect absorption. Analysis in the complex frequency plane}
\label{sec:complex}

\begin{figure}[tbp]
\includegraphics[width=13.9cm]{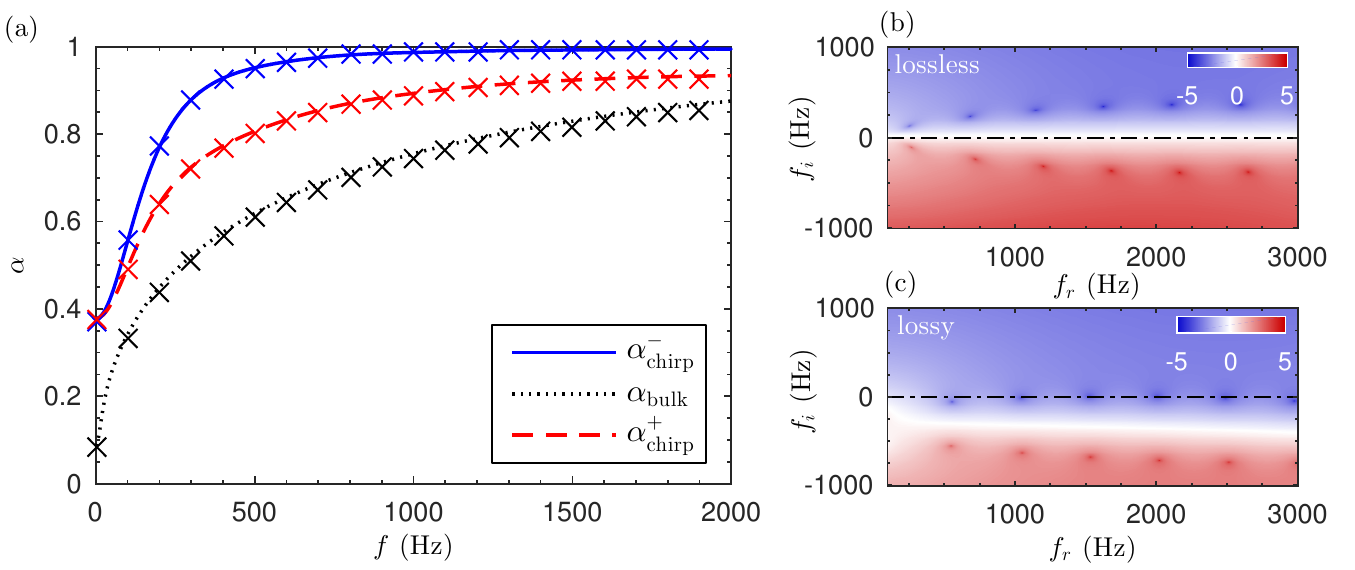}
\caption{Analysis of the absorption coefficient. Blue thick continuous (red dashed) line represents the absorption coefficient of the chirp structure along the (opposite) direction of the chirp. Black thin dotted line represent the absorption coefficient for a bulk porous material with the same dimensions of the chirped multi-layer structure. Markers ($\times$) shows the results obtained by using the FEM. (b-c) Representation of the eigenvalues of the scattering matrix in the complex frequency plane, $\frac{1}{2}\sum_{j=1}^2 \log(|\lambda_j|)$. (b) Artificial lossless case by setting $\mathrm{Im}(K_p)=\mathrm{Im}(\rho_p)=0$. (c) Lossy case.}
\label{fig:fig4}
\end{figure}

The chirped structure is an open cavity that can present two types of losses. Firstly, due to the fact that the system is open, the leakage of energy out of the cavity exists. Secondly, the cavity can have intrinsic losses due to the viscothermal dissipation in the porous materials. Recently in acoustics, the balance of these two kind of losses, known as critical coupling condition, has been exploited to obtain perfect absorption in the subwavelength regime \cite{Romero16a,Romero16b, Merkel15, Jimenez16}. Here, we will use this critical coupling to see that the absorption produced by the chirped is an accumulation of cavity modes. These critically coupled resonances present a small quality factor and therefore overlap to produce a broadband absorption.

We start by analyzing the absorption produced in the system. We study the absorption coefficient from each side of the structure, considering that $\alpha^{(+,-)}=1-|T|^2-|R^{(+,-)}|^2$. Figure~\ref{fig:fig4}~(a) shows the comparison between the absorption coefficient from each side of the chirped structure and that of a bulk porous material with the same thickness as the proposed chirped structure. The absorption of the chirped structure from both sides is always bigger than the absorption of the bulk material due to the fact that the chirped structure presents better impedance matching conditions than the bulk material.

In order to interpret the results concerning the absorption, we analyze the eigenvalues of the scattering matrix in the complex frequency plane, solving the problem for complex frequencies of the form $f=f_r+\imath f_i$. The two eigenvalues are $\lambda_{j}=T\pm\sqrt{R_{\mathrm{chirp}}^+R_{\mathrm{chirp}}^-}$. with $j=1,2$. First, we artificially neglect losses in the system by setting $\mathrm{Im}(K_p)=\mathrm{Im}(\rho_p)=0$. We notice that, due to the properties of the scattering matrix, the poles and zeros of these eigenvalues are symmetric with respect to the real axis, as shown in Fig.~\ref{fig:fig4}~(b). Obviously, in the lossless case the eigenvalues on the real axis are the unity and no absorption is possible. In our case, these poles and zeros are associated to the resonances of the whole system (Fabry-P\'erot resonances). The distance of the poles (zeros) to the real axis in the lossless case is related to the leakage of energy at the resonances. 

Finally, since the losses are introduced, the structure of poles and zeros is shifted to the real axis allowing some zeros eventually cross the real axis \cite{Romero16b} as shown in Fig.~\ref{fig:fig4}~(c).  The case of having a zero in the real axis implies $\sqrt{R_{\mathrm{chirp}}^+R_{\mathrm{chirp}}^-}=T$. In our system, we have previously seen that $T\simeq 0$ for the range of the working frequencies of the chirped structure, therefore, $R_{\mathrm{chirp}}^+R_{\mathrm{chirp}}^-\simeq 0$. Which means that, at the frequencies where the zero is in the real axis, either $R_{\mathrm{chirp}}^+=0$ or $R_{\mathrm{chirp}}^-=0$. Due to the impedance conditions previously discussed, we have seen that $R_{\mathrm{chirp}}^-\simeq 0$ at these frequencies, and as a consequence, $\alpha^-$=1. In this situation the intrinsic losses compensate the leakage of the system and the energy remains trapped in the cavity, i.e., the acoustic energy is completely absorbed being transformed in thermal energy at the resonance frequency.

\section{Conclusions} 
\label{sec:conclusions}         
In this work we have theoretically analyzed the acoustical properties of a chirped multi-layer porous material in transmission. The adiabatic variation of the physical properties inside the structure allows us to characterize the system with local effective parameters, and then to design the system to have particular behavior. In this case, we use the combination of the impedance matching condition with the intrinsic losses of the porous layers to design a broadband unidirectional quasi perfect absorber. Using the formalism of the transfer and scattering matrix method, we have characterized the transmission and reflection properties. Particularly, we have seen, using the eigenvalues of the scattering matrix in the complex frequency plane, how the intrinsic losses can be used to compensate the leakage of the system to produce prefect absorption conditions. The chirped multi-layer porous material analyzed in this work behaves as a resonant cavity, which resonances have a very low quality factor and they overlap to create a broadband unidirectional perfect absorbers. We also show that the chirped structure presents an improvement if compared to a bulky porous structure: using less amount of material the absorption reaches higher values and, in particular, low frequency sound is more efficiently absorbed. Therefore, structures as the one presented here could be of particular interest in civil engineering applications.

\begin{acknowledgments}

This work has been funded by the Metaudible project ANR-13-BS09-0003, co-funded by ANR and FRAE, and also by Ministerio de Econom\'ia y Competitividad (Spain) and European Union FEDER through project FIS2015-65998-C2-2-P.

\end{acknowledgments}

%

\end{document}